# Dynamic nuclear polarization with simultaneous excitation of electronic and nuclear transitions

G. W. Morley[1], K. Porfyrakis[2], A. Ardavan[3] and J. van Tol[4]

1 London Centre for Nanotechnology and Department of Physics and Astronomy, University College London, London WC1H 0AH, UK
2 Department of Materials, University of Oxford, Parks Road, Oxford, OX1 3PH, UK
3 Clarendon Laboratory, University of Oxford, Parks Road, Oxford OX1 3PU, UK
4 National High Magnetic Field Laboratory and Florida State University, Tallahassee Florida 32310 USA

## Abstract

Dynamic nuclear polarization transfers spin polarization from electrons to nuclei. We have achieved this by a new method, simultaneously exciting transitions of electronic and nuclear spins. The efficiency of this technique improves with increasing magnetic field. Experimental results are shown for N@$C_{60}$ with continuous-wave microwaves, which can be expected to produce even higher polarization than the corresponding pulsed techniques for electron spins greater than 1/2. The degree of nuclear polarization in this case can be easily monitored through the intensities of the well resolved hyperfine components in the EPR spectrum. The nuclear spin-lattice relaxation time is orders of magnitude longer than that of the electrons.

## Introduction

Enhanced nuclear polarization is useful in experiments ranging from NMR[1,2] and MRI[3] to neutron scattering[4], high energy physics[5] and quantum computing[6]. This interest has driven the development of a range of different techniques for dynamic nuclear polarization (DNP).

In the most commonly performed experiments, the only step is the application of continuous wave (CW) microwave radiation to excite electronic spin transitions. Three main DNP mechanisms can be distinguished in this case:
- In the *Overhauser effect*[7] polarized electron spins are flipped by electron paramagnetic resonance (EPR) and then 'flip-flop' transitions can occur in which a nuclear spin and an electron spin flip in opposite directions.
- In the *solid effect*[2] microwave radiation is used to drive simultaneous transitions of the electron and nuclear spins.
- The *thermal mixing effect*[2] uses flip-flop transitions between two electron spins and one nuclear spin.

Several extensions of DNP have been demonstrated:
- Experiments in which the sample is suddenly heated from cryogenic temperatures have produced huge DNP enhancements of $\varepsilon > 10000$[1] (where $\varepsilon$ is defined as the nuclear polarization after DNP divided by the nuclear polarization beforehand). However, the constant temperature DNP polarization in these experiments was due to the thermal mixing effect providing a polarization of only 26% at 1.1 K.
- Optical excitation can supply over 60% electronic polarization of pentacene at room temperature leading to DNP enhancements of $\varepsilon=5500$ after 60 minutes of the integrated solid effect[8].
- With pulsed EPR using spin locking pulses around 10 µs long, DNP enhancements of $\varepsilon=10$-14 have been recorded after 20 minutes[9].
- Combining optical excitation with 100 ns EPR pulses has produced DNP with a total irradiation period of just 10 seconds[10].

However, all of the experiments mentioned so far rely on transitions that are not magnetic-dipole allowed, meaning the rates are slower than dipole allowed transitions such as EPR and NMR. As the magnetic field is increased this problem becomes worse, and the DNP rates can be low with respect to the nuclear spin-lattice relaxation rate ($1/T_1^n$), thereby reducing the DNP enhancement.

Any DNP technique at high magnetic fields benefits from a larger electronic polarization for a given temperature. Efficiently using this polarization is the subject of this paper. Further, many applications of DNP require high magnetic fields of 3 – 20 T, where the approaches described so far are not efficient. For example, NMR experiments benefit from higher spectral resolution as the magnetic field is raised.

The problem of non-dipole-allowed transitions can be avoided by driving the electronic transitions with micro/millimetre waves and the nuclear spins with radio-frequency (RF) waves. Electron nuclear double resonance (ENDOR) experiments have been observed to produce nuclear polarization in this way[11].

One experiment has deliberately used NMR to produce DNP, making use of only magnetic-dipole allowed EPR and NMR transitions[12]. However, this used fast passage EPR and did not perform EPR and NMR simultaneously. This technique is not optimum as is discussed below for the different cases in which the electron spin is **S**=1/2 and **S**>1/2.

Materials and Methods

We have increased the nuclear polarization from ENDOR by exciting only those transitions that provide the desired DNP. This technique produces a polarization of nuclear spins enhanced by ENDOR (PONSEE)[6]. PONSEE works best in cases where the EPR spectrum shows well-resolved hyperfine splitting. The EPR of only those spins whose nuclei are in an unwanted spin state are selectively excited. NMR is used to flip those nuclei to the desired spin state. Performing PONSEE with simultaneous EPR and NMR has lead to a nuclear polarization of 62%[6] at a field of 8.6 T and a temperature of 3 K, corresponding to a DNP enhancement factor of $\varepsilon = 1100$. Combining PONSEE with a temperature jump[1] may lead to even larger DNP than either one alone.

The nuclear polarization can be conveniently measured in our experiments simply by comparing the areas of the EPR resonances corresponding to the different nuclear spin states. The EPR is saturated due to the long relaxation times, but the saturation is expected to have the same effect for the different nuclear spin states[13]. Hence the difference in the areas of hyperfine components in our EPR spectrum is attributable to the nuclear polarization. All of the measurements we have performed are consistent with this understanding.

Polarization of nuclear spins enhanced by *pulsed* ENDOR (PONSEPE) would use pulsed electronic and nuclear spin excitation. This is similar to a fast-passage experiment performed previously[12], but PONSEPE could be much faster, particularly when compared to more traditional forms of DNP. If the aim is to enhance nuclear polarization without concern for the electron spin, then it is not necessary to wait for the electronic spin lattice relaxation, $T_1^e$. In this case the PONSEPE sequence consists simply of an electronic $\pi$ pulse followed by a nuclear $\pi$ pulse. The total duration of this preparation would be limited only by that of the nuclear $\pi$ pulse: typically on the order of 20 μs. It may then be possible to create nuclear polarization orders of magnitude faster than previous DNP preparations that take more than 10 seconds.

The source of PONSEPE's speed would be that only a single cycle should be needed. This would avoid the fast-passage scan that takes several seconds[12], and the non-dipole allowed transitions is other DNP schemes. PONSEPE is expected to be most efficient for systems with electon spin **S**=1/2 as discussed below.

N@$C_{60}$ is a nitrogen atom trapped inside a fullerene[14]. The sample was prepared by ion implantation[15] and purified with high performance liquid chromatography (HPLC)[16]. The purified sample was dissolved in molten sulphur in a quartz EPR tube of outer diameter 4 mm. The tube was sealed under a dynamic vacuum, and the sulphur crystallized.

Measurements were made with a high-field CW EPR spectrometer[17] that has recently been upgraded for pulsed operation[18].

Results

Figure 1 shows CW EPR performed with 240 GHz microwave excitation. The $C_{60}$ cage has no EPR signal, but the nitrogen atom has an electronic spin of 3/2. The nuclear spin of nitrogen is **I**=1, apart from an 0.4% abundance of $^{15}$N. Previous PONSEE measurements with $^{15}$N@$C_{60}$ used a small sample leading to low signal to noise ratios[6]. The N@$C_{60}$ sample used for the experiments described here contained orders of magnitude more spins leading to a stronger signal.

The red curve in Fig. 1 shows three resonances with the expected hyperfine splitting of 0.56 mT. The measurement was made with field modulation. The absorption-like lineshape indicates that each resonance is saturated due to the long electron spin-lattice relaxation time[13]. We have previously reported that $T_1^e$ for N@$C_{60}$ in this environment reaches 4.5 minutes at a field of 0.33 T and a temperature of 4 K[19].

While this is a long value for $T_1^e$, it is still much shorter than the $T_1^n$. For any DNP to be effective, the $T_1^n$ should be long compared with the time taken for the nuclear polarization to be generated. For CW PONSEE, this condition corresponds to $T_1^n \gg T_1^e$. The decay of nuclear polarization produced in N@$C_{60}$ by PONSEE indicates that $T_1^n > 1$ day at 4.2 K. This measurement lacks precision because the longest time for which the system was allowed to decay was 38 hours, and only seven such points were recorded.

The CW ENDOR scan in Figure 2 shows the expected resonances at 2.7, 18.5, 34.2 and 50 MHz, corresponding to the nuclear transitions for electron spin $S_z$ = 3/2, 1/2, -1/2 and -3/2 respectively. Each of these resonances consist of the two degenerate transitions $I_z = 0 \leftrightarrow I_z = \pm 1$, as quadrupole effects in this high symmetry system are effectively zero. The ENDOR linewidths were around 50 kHz. A strong artefact is present at 100 MHz (not shown) that is smaller at 50 MHz and smaller still at 25 MHz. We attribute this behaviour to the spectrometer electronics rather than the superimposed ENDOR resonance at 50 MHz.

The black curve in Figure 1 shows the effect of CW PONSEE. This was achieved by holding the magnetic field on the high-field EPR resonance while sweeping the radio frequency (RF) from 1-4 MHz for 45 minutes (though almost the same degree of polarization enhancement was achieved after only 20 minutes).

After PONSEE, the area of the high-field resonance is only 8% of the area of all three resonances. Before PONSEE, the expected thermal population of the high-field resonance is 33.31% with a magnetic field of 8.57 T and a temperature of 4 K. This takes into account the increased hyperfine coupling of $^{14}$N due to its presence in a $C_{60}$ cage[6].

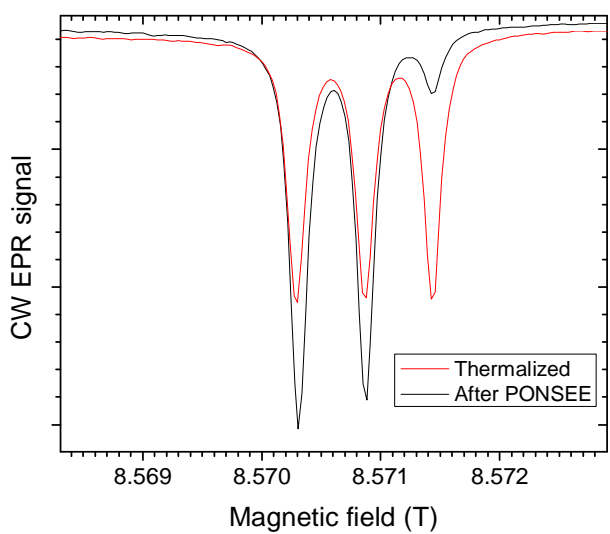

Figure 1. CW EPR measurements before and after CW PONSEE of N@$C_{60}$ using 240 GHz radiation at 4 K.

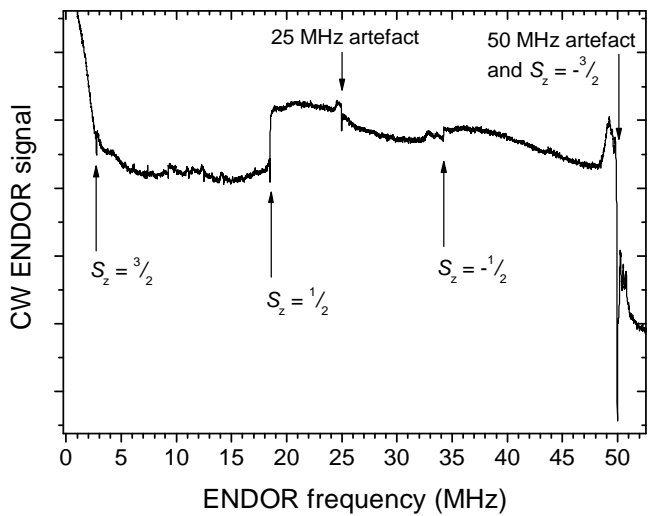

Figure 2. CW ENDOR spectrum of N@$C_{60}$ using 240 GHz radiation at 4 K.

Discussion

For the quantum information applications we are considering[6,19-22], the electron spin should relax back to its polarized state before the preparation is complete. This means that PONSEPE is not much faster than the CW version for these applications, as both processes are limited by the electronic $T_1$. Also it is desired that the states that have not been emptied by PONSEE should contain the same polarization, which is achieved by CW PONSEE, but not with a single cycle of PONSEPE.

For an electron spin **S** > 1/2 (such as the **S** = 3/2 case in N@$C_{60}$), CW PONSEE is expected to provide more total polarization than a single cycle of PONSEPE. Figure 3 shows the reason for this: CW PONSEE provides some nuclear polarization to all molecules, while a single cycle of PONSEPE only affects those molecules that have their electron spin flipped by the π pulse. At the end of a PONSEPE cycle, reapplying steps 2 3 and 4 will lead to an increased nuclear polarization. As the number of cycles is increased the polarization grows towards the CW limit.

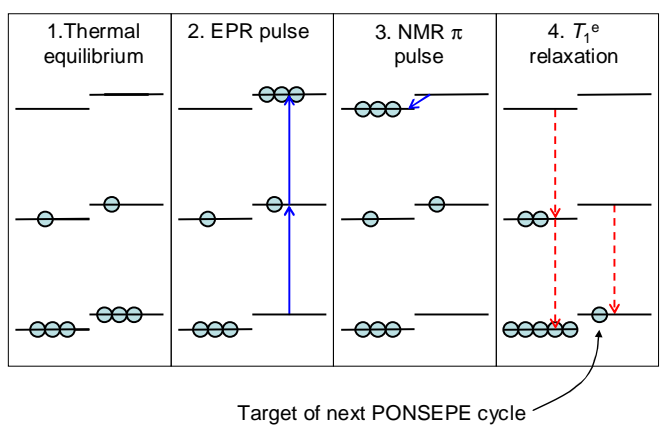

Target of next PONSEPE cycle

Figure 3. Schematic representation of a single cycle of PONSEPE (polarization of nuclear spins enhanced by pulsed ENDOR). In each of the four boxes, the three horizontal lines on the left represent the three electronic spin states with one of the nuclear spin states. The three lines on the right correspond to the same electronic spin states, having the other nuclear spin state. Each ball semi-quantitatively represents some population in an energy level. Electronic spin-lattice relaxation is the only relaxation mechanism considered in this simple picture. A spin system with electron spin **S** = 1 and nuclear spin **I** = 1/2 is chosen as the simplest case in which CW PONSEE provides greater nuclear polarization than a single cycle of PONSEPE.

## Acknowledgements

This work was supported in part by an IHRP grant from the NHMFL and by the EPSRC through the QIP IRC (No. GR/S82176/01) and grant numbers GR/S23506 and EP/D049717. The experiments were performed at the NHMFL (supported by NSF cooperative agreement No. DMR-0084173, by the state of Florida, and by the DOE). A.A. is supported by the Royal Society.